\documentclass[runningheads]{llncs}
\usepackage{graphicx}
\usepackage{xcolor}
\usepackage{amsfonts}
\usepackage{multirow}
\usepackage{amsmath}
\usepackage{comment}

\usepackage{xspace}
\newcommand*{\ipnrn}{NaturalNumber\xspace}
\newcommand*{\fprn}[2]{FixedNaturalNumber(#1, #2)\xspace}
\newcommand*{\ipnrz}{IntegerNumber\xspace}
\newcommand*{\fprz}[2]{FixedIntegerNumber(#1, #2)\xspace}
\newcommand*{\ipnrq}{RationalNumber\xspace}
\newcommand*{\fprq}[3]{FractionalNumber(#1, #2, #3)\xspace}
\newcommand*{\ipnrfixedp}{FixedPoint($\infty$)\xspace}
\newcommand*{\ipnrfloatp}{FloatingPoint($\infty$)\xspace}
\newcommand*{\taperedfloat}[1]{TaperedFloatingPoint(#1)\xspace}
\newcommand*{\floatp}[1]{FloatingPoint(#1)\xspace}
\newcommand*{\fprfixedp}[3]{FixedPoint(#1, #2, #3)\xspace}
\newcommand*{\fprfloatp}[3]{FixedFloatingPoint(#1, #2, #3)\xspace}
\newcommand*{\fprieee}[3]{IEEE754(#1, #2, #3)\xspace}
\newcommand*{\fprmorris}[3]{Morris(#1, #2, #3)\xspace}
\newcommand*{\morrisheb}[3]{MorrisHEB(#1, #2, #3)\xspace}
\newcommand*{\morrisbias}[3]{MorrisBiasHEB(#1, #2, #3)\xspace}
\newcommand*{\morrisunary}[2]{MorrisUnaryHEB(#1, #2)\xspace}
\newcommand*{\fprposit}[3]{Posit(#1, #2, #3)\xspace}

\newcommand*{\red}[1]{\color{red}#1 \color{black}}

\begin{document}
%
\title{A Number Representation Systems Library Supporting New Representations
Based on Morris Tapered Floating-point with \\ Hidden Exponent Bit}

%
%
\author{Ștefan-Dan
Ciocîrlan\inst{1,2}\orcidID{0000-0002-7929-1888} \and \\ Dumitrel
Loghin\inst{2}\orcidID{0000-0002-8965-542X}
\authorrunning{SD. Ciocîrlan et al.}
%
\institute{University Politehnica of Bucharest \and
National University of Singapore} \\
\email{stefan\_dan.ciocirlan@upb.ro} \\
\email{dumitrel@comp.nus.edu.sg}
}
\maketitle              
\begin{abstract}

The introduction of posit reopened the debate about the utility of IEEE754 in
specific domains.
In this context, we propose a high-level language (Scala) library that aims to
reduce the effort of designing and testing new number representation systems
(NRSs).
The library's efficiency is tested with three new NRSs derived from Morris
Tapered Floating-Point by adding a hidden exponent bit. We call these NRSs
MorrisHEB, MorrisBiasHEB, and MorrisUnaryHEB, respectively. We show that they
offer a better dynamic range, better decimal accuracy for unary operations, more
exact results for addition (37.61\% in the case of MorrisUnaryHEB), and better
average decimal accuracy for inexact results on binary operations than posit and
IEEE754. Going through existing benchmarks in the literature, and
favorable/unfavorable examples for IEEE754/posit, we show that these new NRSs
produce similar (less than one decimal accuracy difference) or even better
results than IEEE754 and posit.
Given the entire spectrum of results, there are arguments for MorrisBiasHEB to
be used as a replacement for IEEE754 in general computations. MorrisUnaryHEB has
a more populated ``golden zone'' (+13.6\%) and a better dynamic range (149X)
than posit, making it a candidate for machine learning computations.

\keywords{Number Representation System  \and Tapered Floating Point \and IEEE754 \and Posit \and Computer Arithmetic}
\end{abstract}
%
%
%
\section{Introduction}
Computers are well-known for their ability to run complex mathematical
operations in a very fast way. To run such operations, the operands need to be
represented in practical ways using the finite resources of modern computers.
For example, numbers cannot be represented with infinite precision in computers.
Multiple number representation systems (NRSs) were created to simulate the
infinite world of mathematics. For the real number computations (or rational
numbers, to be more precise) the IEEE754 standard~\cite{8766229} is the norm
from its introduction in 1985. Only recently, Gustafson et
al.~\cite{gustafson2017beating,gustafson2017end} questioned its dominance by
proposing new NRSs such as unum and posit.

The introduction of a new NRS, such as posit, produced multiple research work on
its effect on domains such as scientific computing
\cite{klower2019posits,nolander2021comparative}, artificial intelligence
\cite{johnson2018rethinking,carmichael2019deep,carmichael2019performance,langroudi2018deep,ho2021posit,ho2022qtorch+},
digital signal processing \cite{kant2021implementation},
and computer architecture
\cite{jaiswal2019pacogen,lehoczky2018high,uguen2019evaluating,guntoro2020next,sharma2020clarinet}.
Previously proposed NRSs such as Morris Tapered Floating-Point
\cite{morris1971tapered} and universal real representation  \cite{hamada1983urr}
were revived and re-analyzed. This process of re-analysis and benchmarking is
resource and time-consuming every time a new NRS is introduced.

This article proposes a library with multiple NRS implementations, including
IEEE754, Floating-Point, Morris Tapered Floating-Point, Posit, Rational,
Fractional, and Fixed-Point. In addition, existing benchmarks from the
literature are implemented in this library. The main aim of our library is to
have an easy way for adding a new NRS and test it immediately on already
proposed and well-known benchmarks. This can make the analysis of a new NRS more
efficient and less time-consuming. A secondary aim is to make it easy to add new
benchmarks and make them compatible with all possible NRSs. In this paper, the
library's efficiency and usage are tested by adding three new NRSs derived from
Morris Tapered Floating-point~\cite{morris1971tapered} by using the concept of
hidden exponent bit.

Currently, there are a few libraries that implement NRSs. Some of them support
only one NRS: posit (SoftPosit\footnote{https://gitlab.com/cerlane/SoftPosit},
Posit Mathematica Notebook~\cite{gustafson2017posit}, Posit
Octave\footnote{https://github.com/diegofgcoelho/positsoctave},
Julia\footnote{https://github.com/interplanetary-robot/SigmoidNumbers}),
high-precision floating-point (The GNU Multiple Precision Arithmetic
Library\footnote{https://gmplib.org/}, High Precision Arithmetic
Library\footnote{https://www.nongnu.org/hpalib/},
Flexfloat~\cite{tagliavini2018flexfloat}). FloPoCo~\cite{de2011designing} is a
library with floating-point and posits~\cite{MurilloEtcPosit2020}, but its scope
is to generate arithmetic cores for FPGAs. These libraries do not offer a high
spectrum of changeable attributes and NRSs. There is a need for a library that
lets the developer change attributes like size, exponent size, fraction size,
size of the exponent size, rounding method, rules for underflow and overflow.
The solution is found in Universal Numbers
Library~\cite{Omtzigt:2022,Omtzigt2020} which has multiple NRSs (minus Morris,
plus unum type 1 and 2 and valids), a good set of benchmarks, a better
performance and ways of adding new NRSs. The differences between our NRS library
and Universal Numbers Library are the programming language (Scala vs. C++) and
scope (easy to add and test new NRSs vs. performance for new NRSs). We believe
these two libraries are complementary, not competitors. Our NRS library can be
used for designing and benchmarking new NRSs. The filtered NRSs can then be
implemented for performance in the Universal Numbers Library.

To show the efficiency of our library, we design and evaluate three new NRSs
based on Morris Tapered Floating-point with a hidden exponent bit. These three
proposed NRSs are denoted by \morrisheb{n}{g}{r}, \morrisbias{n}{g}{r}, and
\morrisunary{n}{r}, where $n$ is the size (number of bits), $g$ is a parameter
that dictates the size of the exponent size, and $r$ is the rounding rule. These NRSs
were easily added by using the super-class \taperedfloat{size} and then
implementing their underflow, overflow, exponent, and binary representation
rules without any effort on mathematical operations. They are evaluated under
characteristics, unary operations, binary operations, and literature benchmarks,
with the following results:
\begin{itemize}
     \item Better dynamic range than posit and IEEE754. 
     \item \morrisunary{16}{r} has $13.6\%$ more unique values in the ``golden
     zone'' than \fprposit{16}{2}{r}.
     \item Increased number of unique values compared to the basic Morris tapered
     floating-point \cite{morris1971tapered}.
     \item Better decimal accuracy for unary operations.
     \item More exact results for addition ($37.61\%$ more
     in the case of \morrisunary{12}{RE}).
     \item Better decimal accuracy for inexact values on binary operations.
\end{itemize}

To summarize, we make the following contributions in this paper. First, we
design and implement a Scala library that makes it easy to add, test, and
fine-tune number representation systems (NRSs). The library implements a series
of well-known benchmarks from the literature. Secondly, we introduce three new
NRSs based on Morris tapered floating-point, and thirdly, we analyze these three
proposed NRSs together with well-known NRSs such as IEEE 754 floating-point and
posit.

The remainder of this paper is structured as follows. The second section
contains the motivation behind, the scope of, and the architecture of the
library. In the third section, a brief on NRSs implemented in the library can be
found together with some decisions taken throughout the development. The new
NRSs definitions can be found in the fourth section. In the fifth section, we
present the evaluation, before concluding in the sixth section.


\section{Library}
\begin{figure}[tp]
  \centering
  \includegraphics[width=\columnwidth]{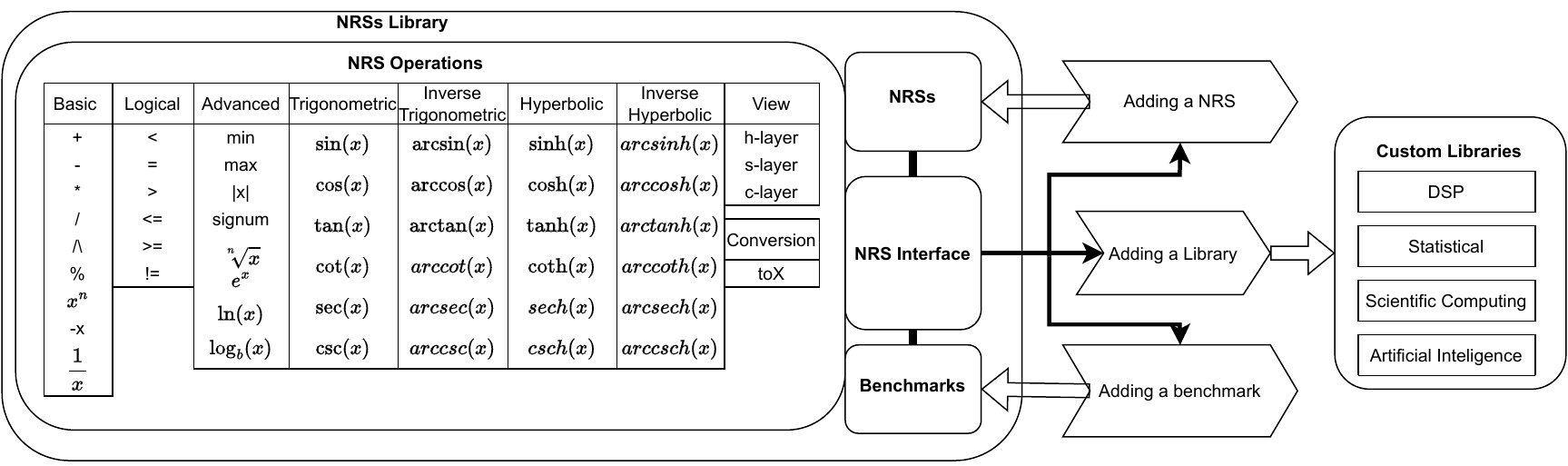}
  \caption{Architecture of the NRS Library}
  \label{fig:arch}
\end{figure}

The main goal of our NRS library is to be an easy-to-use platform for adding and
testing new NRSs and benchmarks. Figure~\ref{fig:arch} offers an overview of our
library. In this library, an NRS is equipped with basic arithmetic operations
(addition, subtraction, multiplication, division, exact division, modulo, power,
negate, inverse), logic operations (less, equal, not equal, greater, greater
or equal, less or equal), advanced arithmetic operations (minimum, maximum,
absolute value, signum, $n^{th}$ root, exponential, natural logarithm, logarithm),
trigonometric functions (sin, cos, tan, cot, sec, csc), inverse trigonometric
functions (arcsin, arccos, arctan, arccot, arcsec, arccsc), hyperbolic functions
(sing, cosh, tanh, coth, sech, csch), inverse hyperbolic functions (arcsinh,
arccoh, arctanh, arccoth, arcsech, arccsh), conversion functions to other NRSs,
and viewing functions. All the above operations are exposed by an NRS interface
part of the library. This interface is inherited by all the NRS implementations.

When adding a new NRS, a developer needs to inherit the NRS interface and
implement all the operations. Some of them have a generic implementation with
Taylor series. Once done, all the benchmarks already implemented by our library
can be run on this new NRS, without writing any additional code.
If the NRS is derived from floating-point or tapered floating-point, the
developer only needs to implement the rules for accepted exponent values,
underflow, overflow, binary representation, and rounding.
For adding a new benchmark, the developer needs to implement the algorithm using
the generic NRS interface and all the past and future NRS implementations will
be able to run it. The rational numbers NRS can be used as a reference, but in
some circumstances, the computation time might be too long given its infinite
precision. In such cases, the fractional numbers NRS is a good alternative.

The benchmark suite implemented in our library contains unary operations, binary
operations, density population of the NRS, and literature benchmarks (as we
shall see in Section~\ref{sec:lit_bench}).
This makes the life of scientists and developers much easier. The scientist will
only focus on developing a new NRS, knowing that many existing benchmarks will
be able to test it without writing additional code. Similarly, when adding a new
benchmark, all the existing NRS implementations will automatically work with it.

Developers of custom libraries can use the NRS interface to implement their
specific functions. There is an opportunity for developing libraries for
statistics, artificial intelligence, or digital signal processing. Currently,
there are some statistics and scientific methods implemented in the library.
For a digital signal processing library, the \textit{complex} construction with
all its operations, FFT and IFFT are already implemented in the NRS Library.
With time, the NRS interface might support more operations, but the current ones
will always remain.

\section{Number Representation Systems}
The library implements the NRSs in Table~\ref{tab:nrss}.
\ipnrn influences all the other NRSs. In the current version of the library,
\ipnrn takes advantage of \texttt{Scala.BigInt}, improving performance and code
readability. \fprn{n}{r} uses \ipnrn to keep the value. Most of the operations
use \ipnrn in background and the result is converted such that it uses exactly
$n$ bits. The results that need more bits are considered Not Representable (NR).
\ipnrz uses \ipnrn to keep the absolute value and a boolean variable for the
sign. For \ipnrz, the euclidean division was chosen because of its mathematical
proprieties. \fprz{n}{r} uses \ipnrz as the value keeper. If the value is using
more bits than the given size, the number becomes NR.

\begin{table}[tp]
\caption{NRSs implement by the library}\label{tab:nrss}
\resizebox{\columnwidth}{!}{%
\begin{tabular}{|l|p{36em}|}
\hline
\multicolumn{1}{|c|}{\textbf{NRS}} & \multicolumn{1}{|c|}{\textbf{Description}} \\
\hline
\hline
\ipnrn & infinite precision without rounding natural number system\\
\hline
\fprn{size}{r} & fixed precision with rounding natural number system, where $size$ is the bit-width and $r$ is the type of rounding\\
\hline
\ipnrz & infinite precision with no rounding sign-magnitude integer number system\\
\hline
\fprz{size}{r} &  fixed precision with rounding integer number system, where $size$ is the bit-width and $r$ is the rounding\\
\hline
\ipnrq & infinite precision with no rounding fractional system\\
\hline
\fprq{n}{m}{r} & fixed precision with rounding fractional system where $n$ is the size of the numerator, $m$ is the size of the denominator, and $r$ is the type of rounding used\\
\hline
\ipnrfixedp & infinite precision with no rounding fixed point system\\
\hline
\fprfixedp{is}{fs}{r} & fixed precision with rounding fixed point system, where $is$ represents the integer size, $fs$ is the binary point value (fraction size), and $r$ is the type of rounding used\\
\hline
\ipnrfloatp & infinite precision with no rounding floating-point number system\\
\hline
\floatp{fs} & floating-point system with infinite precision exponent, no rounding, and with $fs+1$ (fraction size) bits of mantissa (most significant bit always set)\\
\hline
\fprfloatp{es}{fs}{r} & finite precision with rounding floating-point system, where $es$ is the exponent size (exponent in bias form), $fs$ is the fraction size, and $r$ is the type of rounding\\
\hline
\fprieee{es}{fs}{r} & fixed precision with rounding IEEE754 system, where $es$ is the exponent size, $fs$ is the fraction size, and $r$ is the type of rounding\\
\hline
\taperedfloat{size} & tapered floating-point system with infinite precision exponent, no rounding, and with $size$ bits of mantissa\\
\hline
\fprmorris{size}{g}{r} & fixed precision with rounding Morris tapered floating-point NRS, where $size$ is the bit-width, $g$ is the size of the exponent size, and $r$ is the type of rounding\\
\hline
\fprposit{size}{es}{r} & fixed precision with rounding posit NRS, where $size$ is the bit-width, $es$ is the exponent size, and $r$ is the type of rounding\\
\hline
\end{tabular}
}
\end{table}

The problem with the fractional system is that it can overflow easy and there
are multiple ways to represent the same value. The advantage of the fractional
system is that it can represent the entire rational number set $\mathbb{Q}$ in
infinite precision. The overflow problem has a partial solution in doing the
greater common divisor of the numerator and denominator and dividing both by its
value. This solution does not solve the entire problem and adds considerable
computation time. An extension of \fprq{n}{m}{r} proposed to solve the overflow
problem is to divide by two (shifting right by one) both the numerator and
denominator when one of them is out of the given size. \fprq{n}{m}{r} uses
\ipnrq to keep its value.

\textit{Fixed point} representation is similar to \textit{2’s complement}
integer NRS, but it has an attribute called \textit{binary point}. In a simple
way, the value given by the 2’s complement integer NRS is divided by two to the
power of the value of the binary point. This system has an overflow and
underflow problem. Its range of values is smaller than an integer NRS.
\ipnrfixedp keeps its value as a \ipnrq. The only requirement for this NRS is
that the denominator needs to be a power of two. \fprfixedp{is}{fs}{r} takes
another approach by keeping its value as an \ipnrz.

In a floating-point system, a number is represented as $(-1)^{sign} \times
2^{exponent}  \times 1.f$. In infinite precision, $1.f$ can be seen as a
\ipnrfixedp with values inside the $[1,2)$ interval. The first bit represents
the sign of the number, the next $es$ (exponent size) bits represent the
exponent (usually as a bias integer NRS) and the remaining $fs$ (fraction size)
bits represent the fraction bits (usually, there is a hidden bit with the value
$1$ which is the most significant bit). The value is given by $(-1)^{sign}
\times 2^{exponent}  \times (1+\frac{f}{2^{fs}})$, where $f$ represents the
value of the fraction bits without the hidden bit. A problem with the
floating-point system is that $0$ cannot be represented. A solution to this is
that when all bits except the sign bit are $0$, the number is $0$. This creates
the problem of having both $+0$ and $-0$. Given its infinite precision,
\ipnrfloatp does not have the concept of $\infty$.
\floatp{fs} incorporates the concept of $+\infty$ and $-\infty$. Other NRSs are
derived from it, with different rules for the range of the exponent, underflow,
overflow, rounding, and binary representation. One of these NRSs is
\fprfloatp{es}{fs}{r}. \fprfloatp{es}{fs}{r} rules are: (i) in the case of
underflow, the value is round to zero when the exponent is smaller than the
minimum exponent and the rounding rule does not change this, (ii) in the case of
overflow, the value goes to $\infty$ if the exponent value is greater than the
maximum exponent, (iii) the exponent is in bias form. To represent $\infty$, all
the bits except the sign bit are $1$.

The second floating-point system is the standard called
IEEE754~\cite{zuras2008ieee}. This NRS introduces special cases for the smallest
and biggest exponent values. When the exponent has the minimum value, the hidden
bit is zero and the numbers are called subnormals, except for $+0$ and $-0$. The
value of subnormals is given by $(-1)^{sign} \times 2^{\text{-bias}+1} \times
(0+\frac{f}{2^{fs}})$, and the process is called gradual underflow. For maximum
exponent value, different bit strings for fraction and sign bit can represent
\textit{qNaN} (quiet Not A Number -- where the first bit of the fraction is
1), \textit{sNaN} (signal NaN -- where the first bit of the fraction is 0, but
there is another bit in the fraction different from 0), $+\infty$ (all fraction
bits are 0), and $-\infty$ (the same as $+\infty$, only that the sign is
negative).

In our library, \floatp{fs} does not have a binary representation, but it is
used for implementing other NRSs. It is a super-class NRS. It has a boolean
value for the sign, an \ipnrz exponent, a \ipnrn mantissa, some bits used for
rounding and the fraction size value. It implements all the operations with the
scope of having a mantissa in the range $[1,2)$ by making sure that every
operation produces at least $fs + 1$ bits of mantissa with the most significant
one having the value $1$, except for the case when the result is $+0$ or $-0$
and the mantissa is $0$. All the additional bits produced by the operation are
appended to a rest bits list. This super-class is used further by
\fprfloatp{es}{fs}{r} and \fprieee{es}{fs}{r}. \fprfloatp{es}{fs}{r} and
\fprieee{es}{fs}{r} are doing the operations using \floatp{fs} and then
verifying the results with their rules for underflow, overflow, and binary
representation. Specifically, in the case of \fprieee{es}{fs}{r}, if the
exponent is smaller or equal compared to the minimum exponent value the number
might be a subnormal number. This is tested by the difference between the
subnormal exponent and the exponent value. If it is not a subnormal number, it
underflows to $0$.

Beside issues such as multiple representations for zero, subnormal numbers, and
too many bit representations for NaN, IEEE754 might have an oversized or
undersized exponent for a given problem. Morris observed this and introduced
tapered floating-point~\cite{morris1971tapered}, adding an extra field
representing the exponent size. This means that the exponent size and the
fraction size are dynamically computed. A Morris floating-point system is
determined by the bit-width and the size of the exponent size denoted by $g$. The
first $g$ bits represent the $G$ value which is used to compute the exponent
size as $es = G + 1$. The next bit is the exponent sign bit followed by $es$
bits that represent the absolute value of the exponent. The next bit is the
fraction sign, and the remaining bits are considered the fraction bits. The
hidden bit is always $1$. The final value is computed as $2^{\textit{exponent}}
\times (-1)^{\textit{fraction sign}} \times (1+\frac{f}{2^{fs}})$, where
$\textit{exponent}=(-1)^{\textit{exponent sign}} \times
\textit{exponentBinaryValue}$ and \textit{exponentBinaryValue} is a natural
number, not in bias form. Zero is represented by all bits $0$ and error cases
(NaN) are represented when all the bits are $1$. Small numbers have a better
precision because more fraction bits are allocated to represent them.
In most cases, the dynamic range of a Morris NRS is bigger compared to IEEE754
of the same size, for the obvious reason that more bits can be used for the
exponent. However, these NRSs still have the issue of multiple
representations for the same value.

Similar to \floatp{fs}, we created \taperedfloat{size} in our library to help
with implementing all the operations for tapered floating-point NRSs. In this
case, the fraction size is not known so the system uses the bit-width as the
exact fraction size. The fraction is in the range $[1,2)$ and it has a hidden
bit with value $1$. The \taperedfloat{size} NRS does not contain $\infty$.
Every NRS derived from it needs to add rules for underflow, overflow, rounding,
and binary representation.

In trying to solve the problems of current floating-point systems, Gustafson
proposed posit~\cite{gustafson2017beating} (which is another type of
\taperedfloat{size}). A posit NRS is determined by its total size and exponent
size ($es$). The first bit in a posit s-layer representation is the sign bit.
The concept for negative numbers is similar to 2's complement: all the bits are
negated, and one is added to the value. Next, it is the regime field which is
dynamic and uses unary arithmetic representation. If the regime starts with a
bit $1$, then it is a positive regime and the consecutive $1$s are counted until
a $0$ is found or the end of the representation is reached. The value of the
regime is $\text{NoC1} - 1$, where NoC1 is number of consecutive $1$s.
Otherwise, if the regime starts with a $0$, then it is a negative regime and the
consecutive $0$s are counted until the first bit of $1$ is found or the end of
the representation is reached. The regime value in this case is $-1 \times
\textit{NoC0}$, where \textit{NoC0} is the number of consecutive $0$s. After the
regime bits, the next exponent size ($es$) bits represent the exponent value in
a base 2 natural number NRS, and the remaining bits are considered fraction
bits. The hidden bit is always $1$. The final value of a posit is given by
$(-1)^{\textit{sign}} \times 2^{\textit{regime} \times 2^{es} +
\textit{exponentBinaryValue}} \times (1+\frac{f}{2^{fs}})$. There are two
special cases: for zero, when all the bits are $0$, and for $NaR$ (Not a Real --
positive infinity and negative infinity), when first bit is $1$ and the others
are $0$. Posit solves the problem of a value having multiple representations.
This is a strong propriety for an NRS. Like Morris tapered floating-point
system, posit has better precision for small number creating a posit ``golden
zone'' where all the operations have a better accuracy than other NRSs. In our
library, \fprposit{n}{es}{r} underflows to the minimum value that is not $0$ and
overflows to the maximum value but not $\infty$. In contrast,
\fprmorris{n}{g}{r} underflows to $0$ and overflows to NR.



\section{New NRS Based on Moris Tapered Floating-point}
In this section, we introduce the three new representations based on Morris
tapered floating-point with a hidden exponent bit. 

\subsection{\morrisheb{size}{g}{r}}

The tapered floating-point introduced by Morris in  \cite{morris1971tapered}
seems a good concept. Its utilization was shown under the posit system proposed
by Gustafson~\cite{gustafson2017beating}. The major problem is the multiple ways
of representing the same number. A solution for this is in borrowing the concept
of hidden bit from mantissa. The $g$ field not only represents the $G$ value
which dictates the exponent size but also the position of the most significant
bit set in the exponent. If the value of the exponent size is kept as $G+1$,
then the minimum absolute value of the exponent is 2 when $G=0$. The exponent
value is $\textit{exponent} = \textit{exponent sign} \times ( (1 \ll es) +
\textit{binaryExponent} )$. There is a need for having zero as exponent value. A
solution for this is to change the formula for exponent size to $es=G-1$. The
exponent is now:
\begin{equation}
    \text{exponent} = \begin{cases}
    (-1)^{\text{exponent sign}} \times ( 2^{es} + \text{binaryExponent} ), & es \neq -1\\
    0, & es=-1.
  \end{cases}
\end{equation}
This NRS is called \morrisheb{size}{g}{r}. 

The next formula is used for computing the value of all the three new NRSs
binary representations presented in this section:
\begin{equation}
    \text{value} = \begin{cases}
    0, & \text{all bits 0}\\
    \text{NR}, & \text{first bit 1 and the rest 0s} \\
    (-1)^{\text{sign}} \times 2^{\text{exponent}} \times (1 + \frac{\text{f}}{2^{\text{fs}}}), & \text{otherwise} \\
  \end{cases}
\end{equation}
The differences are in the way $es$ and $exponent$ are computed.
\morrisheb{size}{g}{r} underflows to $0$, overflows to NR, and uses
\taperedfloat{size} for implementing the operations. 

The binary representation starts with the sign bit. The next $g$ bits represent
the $G$ value in natural base 2 format. The exponent sign bit follows the $g$
field. The next $e$ bits ($es=G-1$) or the next remaining bits (whichever is
smaller) represent the binary exponent value in natural base 2 format. If $es$
is grater than the remaining bits, the remaining bits represent the most
significant bits of the binary exponent value. Te remaining least significant
bits of the binary exponent value will be considered $0$. After taking the
exponent bits, the remaining bits are fraction bits and their count represents
the fraction size. In summary, the binary format is:
\begin{equation}
    \textcolor{red}{s_{f}}\textcolor{blue}{G\textsubscript{g-1}G\textsubscript{g-2}...G\textsubscript{0}}\textcolor{brown}{s_{e}}\textcolor{brown}{e\textsubscript{es-1}e\textsubscript{es-2}...e\textsubscript{0}}\textcolor{orange}{f\textsubscript{fs-1}f\textsubscript{fs-2}...f\textsubscript{0}}
\end{equation}

\subsection{\morrisbias{size}{g}{r}}

One might argue that the problem of multiple representations is still not solved
because even the exponent may have multiple values (for $es=-1$ the exponent
sign does not matter). The problem stems from having a bit dedicated to the
exponent sign. This is already solved in IEEE754 by using a bias value. A bias
value $g$ is proposed. The exponent sign is the sign of $G$ and the exponent
size is $es=|G|-1$. Another issue with Morris and MorrisHEB representations is
that they do not have an order in binary form. A solution for this is to have
the bits of the exponent negated when $G$ is negative. This makes it easy to
implement a hardware compare unit. The NRS with these features is called
\morrisbias{size}{g}{r}, where the exponent is:
\begin{equation}
    \text{exponent} = \begin{cases}
    \text{signum(G)} \times ( 2^{es} + \text{binaryExponent} ), & es \neq -1\\
    0, & es=-1.
  \end{cases}
\end{equation}

\morrisbias{size}{g}{r} underflows to $0$, overflows to NR, and uses
\taperedfloat{size} for implementing the operations. The binary representation
starts with the sign bit. The next $g$ bits represent the $G$ value in bias
format with $\textit{bias}=2^{g-1}-1$. This means that $G=\textit{binary G} -
\textit{bias}$.
The next $es$ bits ($es=|G|-1$) or the next remaining bits (whichever is
smaller) represent the exponent in natural base 2 format, if the signum(G) is
$1$. Otherwise, they need to be negated an the results is the binary exponent
value. If $es$ is grater than the number of the remaining bits, the remaining
bits represent the most significant bits of the binary exponent value. The
remaining least significant bits of the binary exponent value are considered
$0$. After taking the exponent bits, the remaining bits are fraction bits and
their count is the fraction size. In summary, the binary format is:
\begin{equation}
    \textcolor{red}{s}\textcolor{blue}{G\textsubscript{g-1}G\textsubscript{g-2}...G\textsubscript{0}}\textcolor{brown}{e\textsubscript{es-1}e\textsubscript{es-2}...e\textsubscript{0}}\textcolor{orange}{f\textsubscript{fs-1}f\textsubscript{fs-2}...f\textsubscript{0}}
\end{equation}

\subsection{\morrisunary{size}{r}}

Can \morrisbias{size}{g}{r} be further improved? From the last standard of
posit~\cite{gustafson2017beating}, we are inspired by the choice for fixing the
exponent size to make it dependent only on the size and making the conversion
between different sizes easier. This can be adapted using an unary
representation for the $g$ value (similar to the \textit{regime} in posit).
There is also a need for the exponent size value of $-1$, so the
formula for the exponent size is:
\begin{equation}
    \text{exponent size} = \begin{cases}
    -k - 1, & k < 0 \\
    k - 1, & k \ge 0.
  \end{cases}
\end{equation}
where $k$ is the regime.

\morrisunary{size}{r} underflows to $0$, overflows to NR, and uses
\taperedfloat{size} for implementing its operations. The binary representation
starts with the sign bit. The next bit represents the first regime bit $r_0$.
The next consecutive bits with the same value as $r_0$ are considered regime
bits.
The next bit after them, if it exists, has the negated value of $r_0$ and it is
also considered as part of the regime. The regime $k$ is computed as:
\begin{equation}
    k = \begin{cases}
    -\text{NoC0}, & r_0=0 \\
    \text{NoC1} - 1, & r_0=1.
  \end{cases}
\end{equation}
The next $es$ bits or the next remaining bits (whichever is smaller) represent
the exponent value in natural base 2 format, if the signum(k) is $1$. Otherwise,
they need to be negated and the result is the binary exponent value.
If $es$ is grater than the remaining bits, the remaining bits represent the most
significant bits of the binary exponent value. The remaining least significant
bits of the binary exponent value are considered $0$. The exponent is computed
as:
\begin{equation}
    \text{exponent} = \begin{cases}
    \text{signum(k)} \times ( 2^{es} + \text{binaryExponent} ), & es \neq -1\\
    0, & es=-1.
  \end{cases}
\end{equation}
After taking the exponent bits, the remaining bits are fraction bits and their
count is the fraction size. In summary, the binary format of
\morrisunary{size}{r} is:
\begin{equation}
    \textcolor{red}{s}\textcolor{blue}{r\textsubscript{0}r\textsubscript{1}...r\textsubscript{rs-2}\overline{r\textsubscript{rs-1}}}\textcolor{brown}{e\textsubscript{es-1}e\textsubscript{es-2}...e\textsubscript{0}}\textcolor{orange}{f\textsubscript{fs-1}f\textsubscript{fs-2}...f\textsubscript{0}}
\end{equation}

\section{Evaluation}
In this section, we evaluate the three new proposed NRSs in addition to
well-know NRSs from the literature. In the first subsection, we present the NRSs
under evaluation and their characteristics such as minimum absolute value,
maximum absolute value, dynamic range, and density of numbers in logarithmic
scale. The second subsection presents the decimal accuracy of the unary
operations for the tested NRSs with CDF graphs. In the third subsection, the
color maps of binary operations are presented. The last subsection goes through
some famous literature benchmarks. The next notation are used for rounding in
this section: \textit{RZ} for rounding towards zero and \textit{RE} for rounding
to the nearest tie to even. The values presented in this section are usually
truncated to three decimals after the decimal point.

\subsection{NRSs Under Evaluation and Their Characteristics}

Table~\ref{tab:dynamicrange} presents the NRSs under evaluation with
their minimum absolute value, maximum absolute value, and dynamic range when
the total size is 16 bits. We compare the three new NRSs based on Morris tapered
format with hidden exponent bit with the default Morris representation, fixed
point, fixed floating point, IEEE754, and posit.

\begin{table}[tp]
\caption{16-bit NRSs Dynamic Range}\label{tab:dynamicrange}
\resizebox{\columnwidth}{!}{%
\begin{tabular}{|l|r|r|r|r|r|}
\hline
\multicolumn{1}{|c|}{\textbf{NRS}} & \multicolumn{1}{|c|}{\textit{Min(abs(X))}}
& \multicolumn{1}{|c|}{\textit{Max(abs(X))/$1^{st}$}} & Dynamic Range &
$2^{nd}$ & $3^{rd}$
\\
\hline
\fprfloatp{5}{10}{RE} &  $3.054 \times 10^{-4}$ & $130944$ & $9.6322$ & $130880$ & $130816$\\
\fprfixedp{8}{8}{RE} &  $0.003$ & $127.996$ & $4.515$ & $127.992$ & $127.988$\\
half-IEEE754/\fprieee{5}{10}{RE} &  $5.960 \times 10^{-8}$ & $65504$ & $12.040$ & $65472$ & $65440$\\
\fprposit{16}{2}{RE}  &  $1.387 \times 10^{-17}$ & $72.057 \times 10^{15}$ & $33.715$ & $45.035 \times 10^{14}$ & $11.258  \times 10^{14}$\\
\fprmorris{16}{4}{RZ} &  $9.207 \times 10^{-19710}$ & $1.086 \times 10^{19709}$ & $39418.071$ & $5.887 \times 10^{19689}$ & $3.191 \times 10^{19670}$\\
\morrisheb{16}{4}{RZ}  &  $4.630 \times 10^{-9860}$ & $2.159 \times 10^{9859}$ & $19718.668$ & $3.295 \times 10^{9854}$ & $5.028 \times 10^{9849}$\\
\morrisbias{16}{4}{RE}  &  $6.061 \times 10^{-39}$ & $1.121 \times 10^{77}$ & $115.267$ & $1.085 \times 10^{77}$ & $1.049 \times 10^{77}$\\
\morrisunary{16}{RE} &  $9.168 \times 10^{-2467}$ & $1.090 \times 10^{2466}$ & $4932.075$ & $1.044 \times 10^{1233}$ & $5.809 \times 10^{924}$\\
\hline
\end{tabular}
}
\end{table}

\begin{figure}[tp]
\centering
\includegraphics[width=0.7\columnwidth]{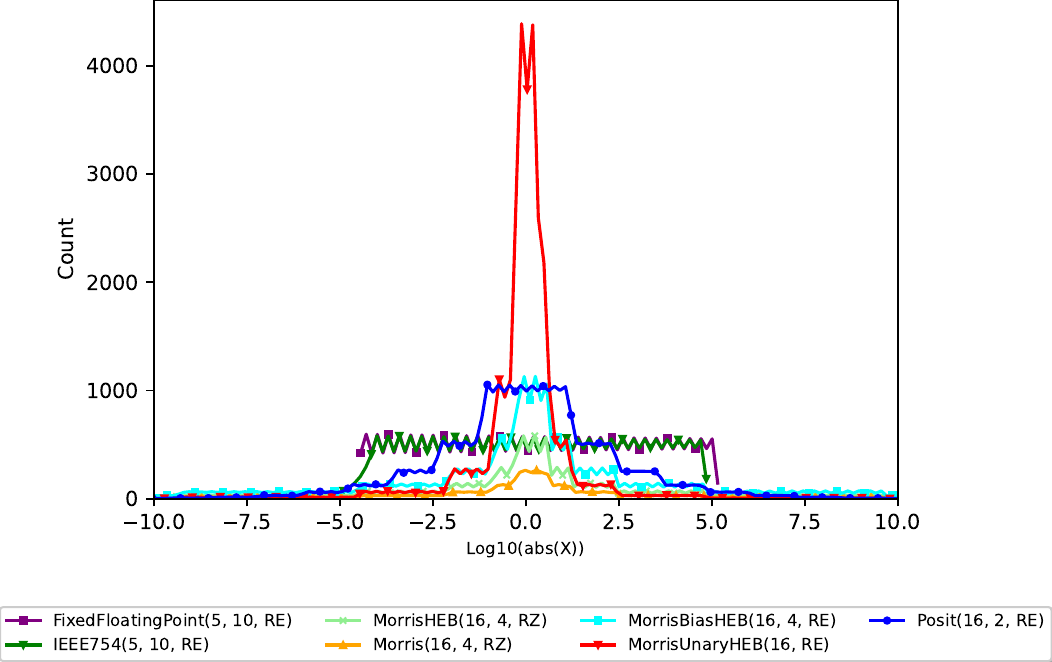}
\caption{Distribution of unique absolute values} \label{fig:densityPlot}
\end{figure}

Tapered floating-point NRSs have a higher dynamic range and can represent higher
and lower absolute values compared to IEEE754 and fixed point. On the other
hand, the difference between consecutive values may be one order of magnitude.
Figure~\ref{fig:densityPlot} presents the count of unique absolute values for
16-bit NRSs on a logarithmic scale. The added value of the hidden exponent bit
can be seen in the increased count of numbers for Morris-derived NRSs. An
interesting result is the \morrisunary{16}{RE} ``golden zone'': it has 30,201
unique absolute values in the interval $(10^{-3}, 10^{3})$ versus 26,587 for
\fprposit{16}{2}{RE}. This, together with the higher dynamic range, makes it a
good competitor for posit in deep neural networks. We shall evaluate this in a
future work.

The difference between the underflow and overflow rules of \fprieee{es}{fs}{r}
and \fprfloatp{es}{fs}{r} can be seen in the gradual underflow for
\fprieee{es}{fs}{r} and the additional higher values for \fprfloatp{es}{fs}{r}.
The usage of a positive regime value for zero can be seen in the unequal
distribution of the values of \morrisunary{16}{RE} and \fprposit{16}{2}{RE} in
Figure~\ref{fig:densityPlot}.

\subsection{Unary Operations}

\begin{figure}[tp]
\centering
\includegraphics[width=0.98\columnwidth]{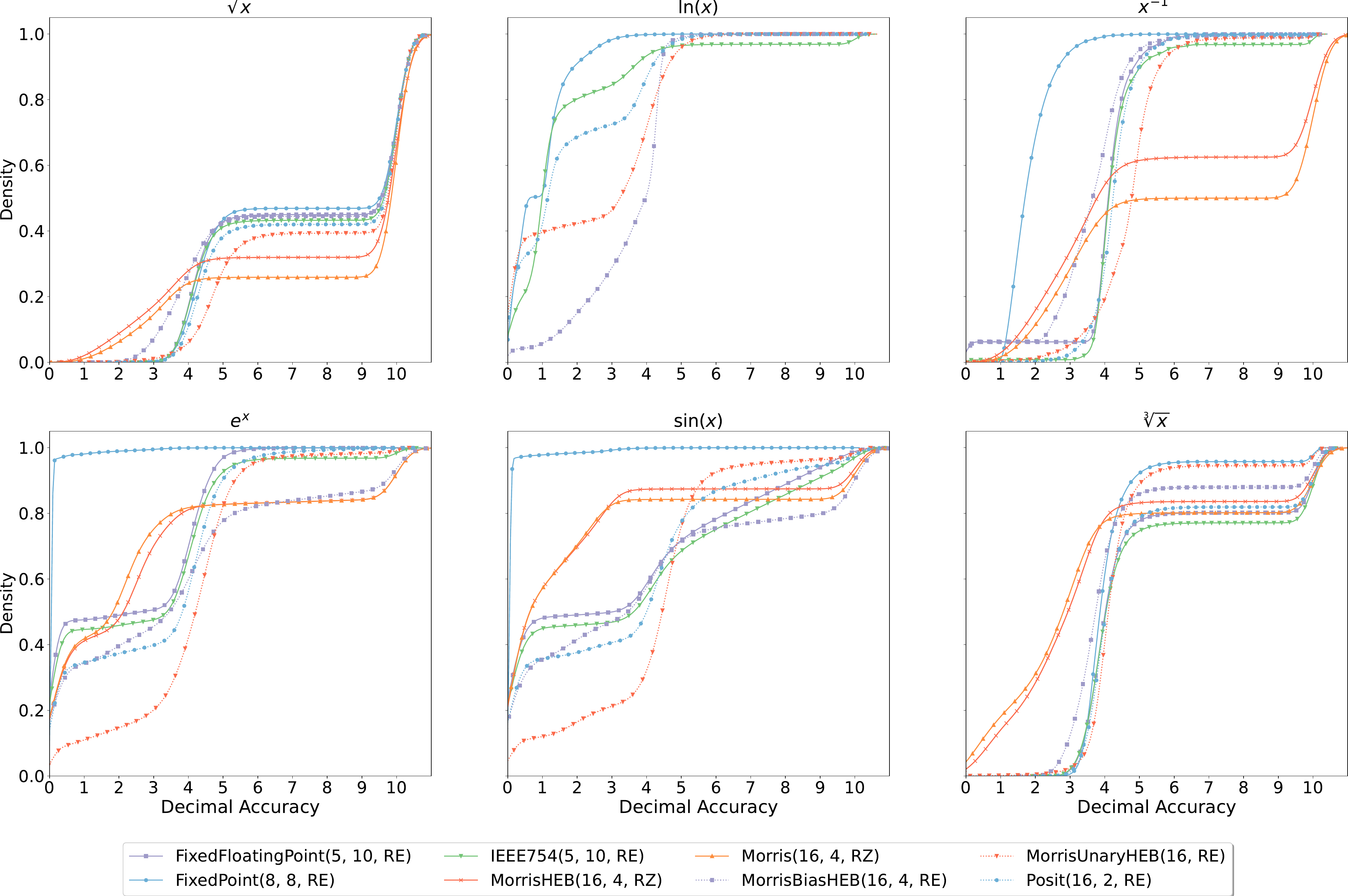}
\caption{CDF of Unary Operations} 
\label{fig:unaryOperations}
\end{figure}

Figure~\ref{fig:unaryOperations} presents the CDF of decimal accuracy for the
square root, natural logarithm, inverse, exponential, sinus, and cube root
operations. The x-axis represents how many accurate digits are there in the
result. For all the Taylor series functions ($\ln(x), \sin(x), e^x$), the
decimal accuracy reference is the \ipnrq result after 30 iterations. For a
decimal accuracy of at least three digits, \morrisunary{16}{RE} is the best NRS.
This is because of its unique absolute values. Note that the exponential is the
only function that increases the magnitude of the result.

\subsection{Binary Operations}

For binary operations, 12 bits NRSs were chosen because 8 bits hold too little
information and 16 bits take too much storage space to keep all the values. We
present the results as color maps, where black represents an accuracy of 10 or
more digits, while white represents zero or less.

\begin{figure}[tp]
\includegraphics[width=\textwidth]{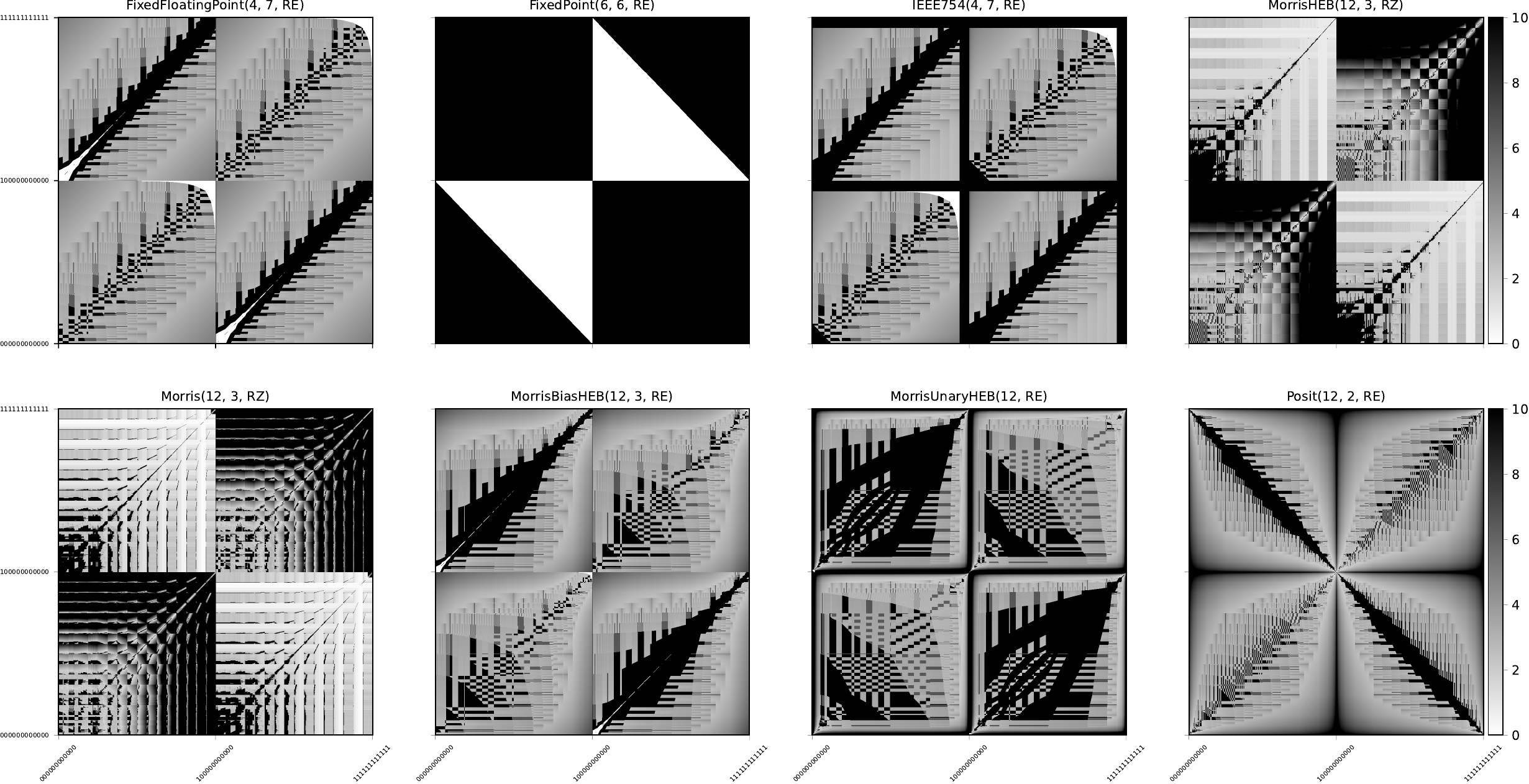}
\caption{Color Maps for Addition} 
\label{fig:addOperation}
\end{figure}


\begin{figure}[tp]
\includegraphics[width=\textwidth]{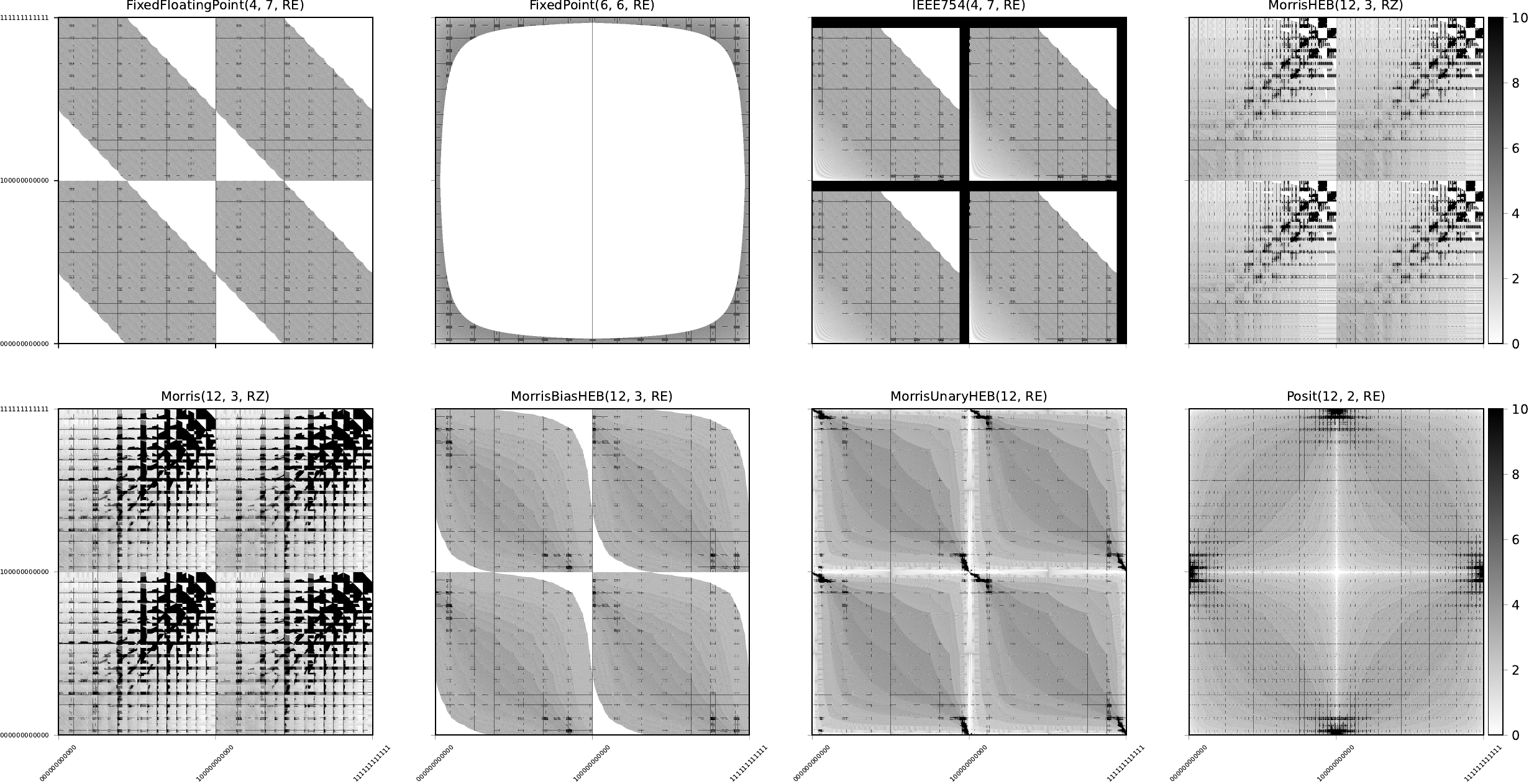}
\caption{Color Maps for Multiplication} 
\label{fig:mulOperation}
\end{figure}

The color map of addition is presented in Figure~\ref{fig:addOperation}. The
subtraction is similar to addition. This plot beautifully shows why
\fprfixedp{is}{es}{r} is the perfect NRS for accumulators if the range of the
results is known. The white color space represents the overflow area for
positive and negative values. The similarities between \fprfloatp{es}{fs}{r} and
\fprieee{es}{fs}{r} are obvious, but one can also observe the effect of the
gradual underflow in \fprieee{es}{fs}{r}. The black border and the plus lines
for \fprieee{es}{fs}{r} represent NaNs (NaN plus anything else results in a
NaN).

The maps for \fprmorris{size}{g}{r} and \morrisheb{size}{g}{r} are different
from the other maps because the binary representations do not represent ordered
values. \morrisbias{size}{g}{r} looks like a mixed between \fprfloatp{es}{fs}{r}
and \fprposit{size}{es}{r}: it exhibits tapered floating-point features by
having an inverse proportional relationship between accuracy and absolute
values. That is, when the absolute values of the operands increase, the decimal
accuracy decreases. \fprposit{size}{es}{r} has a more uniform distribution of
the accuracy. Note that \fprposit{size}{es}{r} does not use sign magnitude but
uses 2's complement for negative numbers so its map symmetry is different from
the other maps.

\begin{table}[tp]
\caption{Binary Operations (ADD,DIV,MUL) Results}\label{tab:binary}
\resizebox{\columnwidth}{!}{%
\begin{tabular}{|l|r|r|r|r|r|r|r|r|r|}
\hline
\multicolumn{1}{|c|}{\multirow{2}{*}{\textbf{NRS}}} & \multicolumn{3}{|c|}{Exact} & \multicolumn{3}{|c|}{Average Accuracy} & \multicolumn{3}{|c|}{Kops}\\
\cline{2-10}
& \multicolumn{1}{|c|}{ADD} & \multicolumn{1}{|c|}{DIV} & \multicolumn{1}{|c|}{MUL} & \multicolumn{1}{|c|}{ADD} & \multicolumn{1}{|c|}{DIV} & \multicolumn{1}{|c|}{MUL} & \multicolumn{1}{|c|}{ADD} & \multicolumn{1}{|c|}{DIV} & \multicolumn{1}{|c|}{MUL} \\
\hline
\fprfloatp{4}{7}{RE} & $16.4\%$ & $2.4\%$ & $2.2\%$ & $3.3$ & $2.4$ & $2.4$ & $191$ & $374$ & $278$\\
\fprfixedp{6}{6}{RE}&  $75.0\%$ & $0.9\%$ & $0.9\%$ & $0.0$ & $2.8$ & $0.5$ & $6535$ & $2551$ & $4117$ \\
\fprieee{4}{7}{RE} ($12.1\%$ NaNs) &  $28.6\%$ &  $14.3\%$ & $14.4\%$ & $3.2$ & $2.7$ & $2.7$ & $362$ & $370$ & $326$ \\
\fprposit{12}{2}{RE} & $12.4\%$ & $4.2\%$ & $4.2\%$ & $2.8$ & $4.0$ & $2.8$ & $150$ & $206$ & $253$ \\
\fprmorris{12}{3}{RZ} & $20.9\%$ & $22.1\%$ & $26.4\%$ & $4.9$ & $1.5$ & $1.5$ & $145$ & $256$ & $347$ \\
\morrisheb{12}{3}{RZ} &  $14.2\%$ & $8.9\%$ & $8.8\%$ & $5.4$ & $1.9$ & $1.8$ & $185$ & $301$ & $385$ \\
\morrisbias{12}{3}{RE} & $20.2\%$ & $2.2\%$ & $2.2\%$ & $3.4$ & $2.7$ & $2.9$ & $148$ & $221$ & $261$\\
\morrisunary{12}{RE} & $37.6\%$ & $1.9\%$ & $1.9\%$ & $4.2$ & $3.0$ & $3.0$ & $142$ & $219$ & $263$ \\
\hline
\end{tabular}%
}
\end{table}

The results of decimal accuracy for multiplication are presented in
Figure~\ref{fig:mulOperation}. The overflow problem of \fprfixedp{is}{es}{r} is
obvious, while gradual underflow helps \fprieee{es}{fs}{r}. The black borders of
\fprieee{es}{fs}{r} are from the NaN values.
\morrisunary{size}{r}, \fprposit{size}{es}{r}, and \morrisbias{size}{g}{r}
exhibit their tapered floating-point proprieties in waves (or bands) of
accuracy. Comparing \morrisunary{size}{r} and \fprposit{size}{es}{r}, the rule
for underflow can be observed as the white band in the color map of
\morrisunary{size}{r}.
These results suggest that the \fprposit{size}{es}{r} rule for underflow might
be the best one to be implemented in an NRS. The results of decimal accuracy for
division are similar to the ones for multiplication and are omitted due to space
constraints.

Table~\ref{tab:binary} presents the percentage of exact results, the average
decimal accuracy for inexact results, and the number of (thousands) operations
per second (Kops).
From \fprieee{4}{7}{RE}, one should remove $12.1\%$ of the results because they
represent NaNs. The interesting results in Table~\ref{tab:binary} are: (i) the
high number of exact results for \morrisunary{12}{RE}, (ii) the relatively good
average decimal accuracy on inexact results for \morrisunary{12}{RE} and
\morrisbias{12}{3}{RE} on all operations, and (iii) the relatively low
percentage of exact results for \fprposit{12}{2}{RE} (this is because of the
increased exponent size).

\subsection{Literature Benchmarks}
\label{sec:lit_bench}

In Table~\ref{tab:enderror}, we summarize the results of the evaluations
proposed by Gustafson in \cite{gustafson2017end}. The proposed evaluations are:
\begin{itemize}
  \item John Wallis Product: $2 \times \prod_{i=1}^{n}\frac{(2 \times i)^2}{(2
\times i - 1) \times (2 \times i + 1)}$ for $n=30$, 
  \item Kahan series: $u_{i+2}=111-\frac{1130}{u_{i+1}}+\frac{3000}{u_{i} \times u_{i+1}}$ for $u_{30}$, 
  \item Jean Micheal Muller: $E(0)=1, E(z)=\frac{e^{z}-1}{z}, Q(x)=|x - \sqrt{x^2 + 1}| - \frac{1}{x + \sqrt{x^2 + 1}}, H(x)=E((Q(x))^2) $ for $H(15),H(16),H(17),H(9999)$, 
  \item Siegfried Rump: $333.74 \times y^6 + x^2 \times (11 \times x^2 \times y^2 - y^6 - 121 \times y^4 - 2) + 5.5 \times y^8 + \frac{x}{2 \times y}$ for $x=77517$ and $y=33096$, 
  \item Decimal accuracy for $r_1$ from Quadratic formula for $a=3,b=100,c=2$,
  \item David Bailey's system of equations: $0.25510582 \times x + 0.52746197 \times y = 0.79981812, 0.80143857 \times x + 1.65707065 \times y = 2.51270273$ solved with Cramer's rule.
\end{itemize}
Note that none of the NRSs passes all the evaluations. The problem is with the limitations of finite representations.

\begin{table}[tp]
\centering
\caption{Benchmarks in \cite{gustafson2017end}}\label{tab:enderror}
\resizebox{0.99\columnwidth}{!}{
\begin{tabular}{|l|r|r|r|c|r|r|}
\hline
\multicolumn{1}{|c|}{\textbf{NRS}} & \multicolumn{1}{|c|}{John Wallis} & \multicolumn{1}{|c|}{Kahan $u_{30}$} & \multicolumn{1}{|c|}{Jean Micheal Muller} & \multicolumn{1}{|c|}{Siegfried Rump} & \multicolumn{1}{|c|}{$r_1$ DA} & \multicolumn{1}{|c|}{David Bailey} \\
\hline
\multicolumn{7}{|c|}{32-bit NRSs} \\
\hline
\fprfloatp{8}{23}{RE} & $3.091$ & $100$ & $(0, 0, 0, 0 )$ & $-63.382 \times 10^{28}$ & $5.612$  &  $(NR, NR)$ \\
\fprfixedp{16}{16}{RE}&  $3.091$ & $100$ & $(1, 1, 1, NR )$ & $NR$ & $3.787$  &  $(NR, NR)$ \\
\fprieee{8}{23}{RE} &  $3.091$ & $100$ & $(0, 0, 0, 0 )$ & $-1.901 \times 10^{30}$ & $5.612$  &  $(NR, NR)$ \\
\fprposit{32}{2}{RE} & $3.091$ & $100$ & $(0, 0, 0, 0 )$ & $1.172$ & $5.996$  &  $(-4, 2)$ \\
\fprmorris{32}{4}{RZ} & $3.091$ & $99.999$ & $(0, 0, 0, 0.995)$ & $20.282 \times 10^{30}$ & $4.599$  &  $(0, 1)$ \\
\morrisheb{32}{4}{RZ} &  $3.091$ & $99.999$ & $(0, 0, 0, 0.989)$ & $15.211 \times 10^{30}$ & $4.945$  &  $(2, 1)$ \\
\morrisbias{32}{4}{RE} & $3.091$ & $100$ & $(0, 0, 0, 0 )$ & $-25.353 \times 10^{29}$ & $5.612$  &  $(1, 0.5)$ \\
\morrisunary{32}{RE} & $3.091$ & $100$ & $(0, 0, 0, 0.999)$ & $1.172$ & $5.612$  &  $(2, 0)$ \\
\hline
\ipnrq & $3.091$ & $6.004$ & $(1, 1, 1, 1)$ & $-0.827$ & $1$  &  $(-1, 2)$ \\
\hline
\end{tabular}
}
\vspace{-10pt}
\end{table}

In Table~\ref{tab:others}, we present the results of multiple benchmarks from
the literature~\cite{de2019posits,goldberg1991every,gustafson2017beating}. These
benchmarks are:
\begin{itemize}
  \item thin triangle area for $a=7,c=b=\frac{7+2^{-25}}{2}$, 
  \item the formula $x=(\frac{27/10 - e}{\pi - (\sqrt{2} + \sqrt{3})})^{67/16}$, 
  \item the fraction $\frac{x^n}{n!}$ for $x=7, n=20$ and $x=25, n=30$, 
  \item Planck constant $h=6.626070150 \times 10^{-34}$, 
  \item Avogadro number $L=6.02214076 \times 10^{23}$, 
  \item speed of light $c=299792458$, 
  \item charge of $\overline{e} 1.602176634 \times 10^{-19}$, 
  \item Boltzmann constant $k=1.380649 \times 10^{-23}$. 
\end{itemize}

The values in Table~\ref{tab:others} represent the decimal accuracy of the
results compared to the correct result. The first two benchmarks are favorable
to \fprposit{size}{es}{r} while the last six are favorable to
\fprieee{es}{fs}{r}.
Morris and its derived NRSs exhibit results that are close to the best NRS for
each benchmark. \morrisbias{size}{g}{r} has good results for the entire spectrum
of benchmarks.

\begin{table}[tp]
\centering
\caption{Other Literature Benchmarks~\cite{de2019posits,goldberg1991every,gustafson2017beating}}\label{tab:others}
\resizebox{0.8\columnwidth}{!}{%
\begin{tabular}{|l|r|r|r|r|r|r|r|r|}
\hline
\multicolumn{1}{|c|}{\textbf{NRS}} & \multicolumn{1}{|c|}{Thin Triangle} & \multicolumn{1}{|c|}{$x$} & \multicolumn{1}{|c|}{$\frac{x^n}{n!}$} & \multicolumn{1}{|c|}{Planck} & \multicolumn{1}{|c|}{L} & \multicolumn{1}{|c|}{c} & \multicolumn{1}{|c|}{$\overline{e}$} & \multicolumn{1}{|c|}{k} \\
\hline
\multicolumn{9}{|c|}{32-bit NRSs} \\
\hline
\fprfixedp{16}{16}{RE}&  $0$ & $2.289$ & $(0, 0)$ & $0$ & $0$  &  $0$  & $0$ & $0$\\
\fprieee{8}{23}{RE} &  $0$ & $4.370$ & $(7.135, 0)$ & $8.727$ & $8.075$  &  $7.839$  & $8.004$ & $7.782$\\
\fprposit{32}{2}{RE} & $1.204$ & $5.684$ & $(4.339, 0)$ & $0.627$ & $4.091$  &  $6.969$ & $4.213$ & $4.037$\\
\fprmorris{32}{4}{RZ} & $0$ & $5.101$ & $(6.016, 5.604)$ & $6.347$ & $6.429$  &  $6.969$ & $7.347$ & $6.480$\\
\morrisheb{32}{4}{RZ} &  $0$ & $5.098$ & $(6.245, 6.188)$ & $6.680$ & $6.784$  &  $7.839$ & $7.347$ & $6.480$\\
\morrisbias{32}{4}{RE} & $0$ & $5.682$ & $(6.911, 8.619)$ & $7.053$ & $7.219$  &  $7.839$ & $7.347$ & $6.878$\\
\morrisunary{32}{RE} & $0$ & $6.875$ & $(6.017, 5.289)$ & $6.031$ & $5.919$  &  $6.969$ & $7.347$ & $5.566$\\
\hline

\end{tabular}%
}
\end{table}


\section{Conclusion}
In this paper, (i) we presented a Scala library that makes it easy to add, test,
and fine-tune number representation systems (NRSs), (ii) we introduced three new
NRSs based on Morris tapered floating-point, and (iii) we analyzed these three
proposed NRSs together with well-known NRSs such as IEEE 754 floating-point and
posit.

By adding the hidden exponent bit to Morris tapered floating-point in three
different forms, the resulting NRSs became competitors for IEEE754 and posit.
\morrisbias{size}{g}{r} exhibits the best results on literature benchmarks on 32
and 64 bits when compared to the other NRSs. On the other hand,
\morrisunary{size}{r} is a great candidate for machine learning computations due
to its ``golden zone'' population, dynamic range, percent of exact results on
addition and average decimal accuracy for inexact results on multiplication.

Our library exhibits a performance of around ~200~Kops which is good enough for
testing and evaluating NRSs, but not enough for real-world applications. In
future works, the library will be integrated with the Aparapi
library\footnote{https://aparapi.com/} and tested on GPU, and used for machine
learning models with Spark. We also plan to increase the number of benchmarks.


\section*{Acknowledgment} Ștefan-Dan Ciocîrlan is partly supported by the
Bitdefender's University PhD Grants Program 2019-2022 and by the Google
IoT/Wearables Student Grants 2022. Dumitrel Loghin is partly supported by the
Ministry of Education of Singapore's Academic Research Fund Tier 1 (grant
251RES2106).

\bibliographystyle{splncs04}


\end{document}